\documentclass[prl,reprint,superscriptaddress,preprintnumbers,nofootinbib,sort&compress]{revtex4-1}
\usepackage{graphicx}
\usepackage{dcolumn}
\usepackage{bm}
\usepackage{color}
\usepackage{amsmath}
\usepackage{amssymb}
\graphicspath{{Figures/}{}{.}}
\usepackage{verbatim,multirow,color}
\usepackage{subfig}
\usepackage{subfloat}

\def\araa{Annu. Rev. Astron. Astrophys }                                  
\def\apj{Astrophys. J. }                                                    
\def\apjs{Astrophys. J. Suppl. Ser. }                                       
\def\apjl{Astrophys. J. Lett. }                                             
\def\aap{Astron. Astrophys. }
\def\aar{Astron. Astrophys. Rev. }                                        
\def\epja{Eur. Phys. J. A }                                               
\def\mnras{Mon. Not. R. Astron. Soc. }                                    
\def\nima{Nucl. Instr. Meth. A }                                          
\def\npa{Nucl.~Phys.~A }                                                  
\def\prc{Phys. Rev. C }                                                   

\begin{document}
\title{Three new low-energy resonances in the $^{22}$Ne(p,$\gamma$)$^{23}$Na reaction}

\author{F.~Cavanna}\affiliation{Universit\`a degli Studi di Genova and INFN, Sezione di Genova, Via Dodecaneso 33, 16146 Genova, Italy}
\author{R.~Depalo}\affiliation{Universit\`a degli Studi di Padova and INFN, Sezione di Padova, Via F. Marzolo 8, 35131 Padova, Italy}
\author{M.~Aliotta}\affiliation{SUPA, School of Physics and Astronomy, University of Edinburgh, EH9 3FD Edinburgh, United Kingdom}
\author{M.~Anders}\affiliation{Helmholtz-Zentrum Dresden-Rossendorf, Bautzner Landstr. 400, 01328 Dresden, Germany}\affiliation{Technische Universit\"at Dresden, Institut f\"ur Kern- und Teilchenphysik, Zellescher Weg 19, 01069 Dresden, Germany}
\author{D.~Bemmerer}\email[e-mail address:~]{d.bemmerer@hzdr.de}\affiliation{Helmholtz-Zentrum Dresden-Rossendorf, Bautzner Landstr. 400, 01328 Dresden, Germany}
\author{A.~Best}\affiliation{Laboratori Nazionali del Gran Sasso (LNGS), 67100 Assergi (AQ), Italy}
\author{A.~B\"oltzig}\affiliation{Gran Sasso Science Institute, 67100 L’Aquila, Italy}
\author{C.~Broggini}\affiliation{INFN, Sezione di Padova, Via F. Marzolo 8, 35131 Padova, Italy}
\author{C.G.~Bruno}\affiliation{SUPA, School of Physics and Astronomy, University of Edinburgh, EH9 3FD Edinburgh, United Kingdom}
\author{A.~Caciolli}\affiliation{Universit\`a degli Studi di Padova and INFN, Sezione di Padova, Via F. Marzolo 8, 35131 Padova, Italy}
\author{P.~Corvisiero}\affiliation{Universit\`a degli Studi di Genova and INFN, Sezione di Genova, Via Dodecaneso 33, 16146 Genova, Italy}
\author{T.~Davinson}\affiliation{SUPA, School of Physics and Astronomy, University of Edinburgh, EH9 3JZ Edinburgh, United Kingdom}
\author{A.~di~Leva}\affiliation{Universit\`a di Napoli Federico II and INFN, Sezione di Napoli, 80126 Napoli, Italy}
\author{Z.~Elekes}\affiliation{Institute for Nuclear Research (MTA ATOMKI), PO Box 51, HU-4001 Debrecen, Hungary}
\author{F.~Ferraro}\affiliation{Universit\`a degli Studi di Genova and INFN, Sezione di Genova, Via Dodecaneso 33, 16146 Genova, Italy}
\author{A.~Formicola}\affiliation{Laboratori Nazionali del Gran Sasso (LNGS), 67100 Assergi (AQ), Italy}
\author{Zs.~F\"ul\"op}\affiliation{Institute for Nuclear Research (MTA ATOMKI), PO Box 51, HU-4001 Debrecen, Hungary}
\author{G.~Gervino}\affiliation{Universit\`a degli Studi di Torino and INFN, Sezione di Torino, Via P. Giuria 1, 10125 Torino, Italy}
\author{A.~Guglielmetti}\affiliation{Universit\`a degli Studi di Milano and INFN, Sezione di Milano, Via G. Celoria 16, 20133 Milano, Italy}
\author{C.~Gustavino}\affiliation{INFN, Sezione di Roma La Sapienza, Piazzale A. Moro 2, 00185 Roma, Italy}
\author{Gy.~Gy\"urky}\affiliation{Institute for Nuclear Research (MTA ATOMKI), PO Box 51, HU-4001 Debrecen, Hungary}
\author{G.~Imbriani}\affiliation{Universit\`a di Napoli Federico II and INFN, Sezione di Napoli, 80126 Napoli, Italy}
\author{M.~Junker}\affiliation{Laboratori Nazionali del Gran Sasso (LNGS), 67100 Assergi (AQ), Italy}
\author{R.~Menegazzo}\affiliation{INFN, Sezione di Padova, Via F. Marzolo 8, 35131 Padova, Italy}
\author{V.~Mossa}\affiliation{Universit\`a degli Studi di Bari and INFN, Sezione di Bari, 70125 Bari, Italy}
\author{F. R.~Pantaleo}\affiliation{Universit\`a degli Studi di Bari and INFN, Sezione di Bari, 70125 Bari, Italy}
\author{P.~Prati}\affiliation{Universit\`a degli Studi di Genova and INFN, Sezione di Genova, Via Dodecaneso 33, 16146 Genova, Italy}
\author{D. A.~Scott}\affiliation{SUPA, School of Physics and Astronomy, University of Edinburgh, EH9 3JZ Edinburgh, United Kingdom}
\author{E.~Somorjai}\affiliation{Institute for Nuclear Research (MTA ATOMKI), PO Box 51, HU-4001 Debrecen, Hungary}
\author{O.~Straniero}\affiliation{Osservatorio Astronomico di Collurania, Teramo, and INFN, Sezione di Napoli, Napoli, Italy}
\author{F.~Strieder}\thanks{Present address: South Dakota School of Mines and Technology, Rapid City, SD 57701, USA}\affiliation{Institut f\"ur Experimentalphysik III, Ruhr-Universit\"at Bochum, 44780 Bochum, Germany}
\author{T.~Sz\"ucs}\affiliation{Helmholtz-Zentrum Dresden-Rossendorf, Bautzner Landstr. 400, 01328 Dresden, Germany}
\author{M.~P.~Tak\'acs}\affiliation{Helmholtz-Zentrum Dresden-Rossendorf, Bautzner Landstr. 400, 01328 Dresden, Germany}\affiliation{Technische Universit\"at Dresden, Institut f\"ur Kern- und Teilchenphysik, Zellescher Weg 19, 01069 Dresden, Germany}
\author{D.~Trezzi}\affiliation{Universit\`a degli Studi di Milano and INFN, Sezione di Milano, Via G. Celoria 16, 20133 Milano, Italy}
\collaboration{The LUNA Collaboration}\noaffiliation

\begin{abstract}
The $^{22}$Ne(p,$\gamma$)$^{23}$Na reaction takes part in the neon-sodium cycle of hydrogen burning. This cycle affects the synthesis of the elements between $^{20}$Ne and $^{27}$Al in asymptotic giant branch stars and novae. 
The $^{22}$Ne(p,$\gamma$)$^{23}$Na reaction rate is very uncertain because of a large number of unobserved resonances lying in the Gamow window. At proton energies below 400\,keV, only upper limits exist in the literature for the resonance strengths. Previous reaction rate evaluations differ by large factors. 
In the present work, the first direct observations of the $^{22}$Ne(p,$\gamma$)$^{23}$Na resonances at 156.2, 189.5, and 259.7\,keV are reported. Their resonance strengths have been derived with 2-7\% uncertainty. In addition, upper limits for three other resonances have been greatly reduced. Data were taken using a windowless $^{22}$Ne gas target and high-purity germanium detectors at the Laboratory for Underground Nuclear Astrophysics in the Gran Sasso laboratory of the National Institute for Nuclear Physics, Italy, taking advantage of the ultra-low background observed deep underground. 
The new reaction rate is a factor of 5 higher than the recent evaluation at temperatures relevant to novae and asymptotic giant branch stars nucleosynthesis. 

\end{abstract}


\maketitle

Recent studies of globular clusters with high-resolution spectrometers \cite{Carretta:2011-AA, Lind:2011-AA, Yong:2003-AA} have opened new windows on galactic chemical evolution. It was found that globular clusters are made up of multiple generations of stars \cite{Gratton12-AAR,Cordero15-AJ}. A peculiar observation in this framework is the anticorrelation between oxygen and sodium abundances found in giant stars belonging to all globular clusters studied so far \cite{Gratton04-ARAA,Carretta09-AA,Johnson12-ApJL}.
The stellar sources responsible for these effects have not yet been identified. One possible source are massive Asymptotic Giant Branch stars (AGB stars with mass $M$\,$>$\,4\,$M_{\odot}$)\footnote{$M_{\odot}$ is the mass of our Sun.} where the so-called Hot Bottom Burning (HBB) process is active \cite{Renzini81-AA,Gratton04-ARAA,Ventura08-AAp}. 
Other possibilities include massive binaries \cite{deMink09-aap}, fast rotating massive stars \cite{Decressin07-aap}, and supermassive (M\,$\sim$\,10$^{4}$\,M$_{\odot}$) stars \cite{Denissenkov14-MNRAS}.

The neon-sodium (NeNa) and magnesium-aluminum (MgAl) cycles of hydrogen burning are activated when the temperature in the hydrogen burning region exceeds $\sim$\,0.07\,GK (corresponding to a Gamow energy of $E_{\rm p}$\,$>$\,100\,keV)\footnote{$E_{\rm p}$ denotes the proton beam energy in the laboratory system.}. In the HBB process, the temperature at the base of the convective envelope rises as high as 0.1\,GK ($E_{\rm p}$\,$\sim$\,120\,keV). The NeNa and MgAl chains increase the surface sodium and aluminum abundances and decrease the magnesium abundance \cite{Denisenkov90-SAL}. In parallel, oxygen is depleted by the oxygen-nitrogen (ON) cycle. The interpretation of the observed abundance patterns thus requires a precise knowledge of nucleosynthesis in the NeNa and MgAl chains and in the ON cycle. The MgAl chain \cite{Strieder12-PLB} and ON cycle \cite{Scott12-PRL} have recently been addressed in low-energy experiments. The present work reports on a high-luminosity experiment on the most uncertain reaction of the NeNa cycle, $^{22}$Ne(p,$\gamma$)$^{23}$Na. 

\begin{figure}[tb]
\includegraphics[width=\columnwidth,trim=4cm 2cm 3cm 0cm]{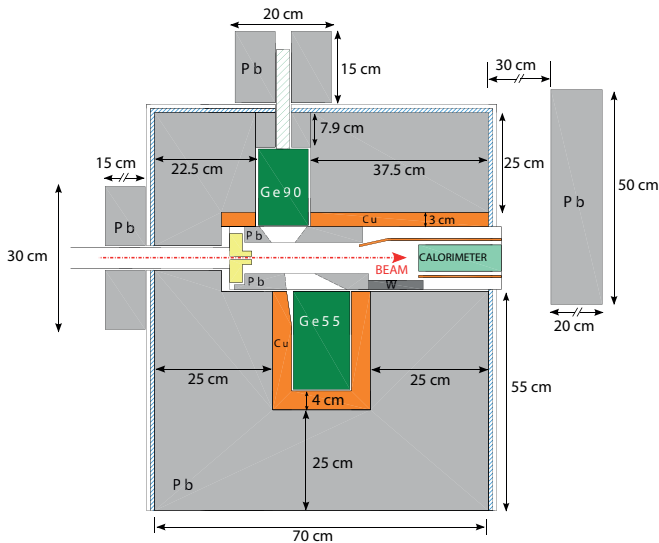}
\caption{Target chamber, germanium detectors (Ge55 and Ge90), copper (orange) and lead (grey) shielding. The  external lead wall on the right hand side covers the shielding gap where the calorimeter is inserted \cite{Cavanna:2014lia,Depalo15-Diss,Cavanna15-Diss}.}
\label{fig:setup}
\end{figure}

In addition to the AGB star scenario and the HBB process, hydrogen burning of $^{22}$Ne also plays a role in explosive nucleosynthesis scenarios. In classical novae (0.15 $<T<$ 0.45\,GK, 150 $<E_{\rm p}<$ 300\,keV, \cite{Jose98-ApJ}), the ejected material carries the products of the hot CNO cycle and of the NeNa \cite{Sallaska10-PRL} and MgAl chains.  For an oxygen-neon nova, the uncertainty on the $^{22}$Ne(p,$\gamma$)$^{23}$Na reaction rate leads to six orders of magnitude uncertainty on the $^{22}$Ne yield \cite{Iliadis02-ApJSS}. For a carbon-oxygen nova, $^{22}$Ne(p,$\gamma$)$^{23}$Na was found to affect the abundances of elements between neon and aluminum \cite{Iliadis02-ApJSS}. As a consequence, there is a call for a more precise $^{22}$Ne(p,$\gamma$)$^{23}$Na thermonuclear reaction rate \cite{Denissenkov14-MNRAS-442}.
In type Ia supernovae, during pre-explosion hydrogen burning ($T<$ 0.6\,GK, $E_{\rm p}<$\,400\,keV) on the surface of the white dwarf star, the $^{22}$Ne(p,$\gamma$)$^{23}$Na reaction may deplete $^{22}$Ne, hence changing the electron fraction and all the subsequent nucleosynthesis \cite{Chamulak:2008-ApJ}.
In core collapse supernova precursors, proton capture on $^{22}$Ne competes with the neutron source reaction $^{22}$Ne($\alpha$,n)$^{25}$Mg, thus affecting neutron capture nucleosynthesis \cite{Pignatari15-ApJL}. Summarizing, new $^{22}$Ne(p,$\gamma$)$^{23}$Na data are needed for several highly topical astrophysical scenarios ranging from AGB stars to supernovae. 

$^{22}$Ne(p,$\gamma$)$^{23}$Na resonances with resonance energy $E_{\rm p}^{\rm res}$\,$>$\,400\,keV affect the thermonuclear reaction rate for high temperatures $T$\,$>$\,0.5\,GK, see \cite{Depalo15-PRC} for recent new data. For lower temperatures $T$\,$<$\,0.5\,GK relevant to most of the scenarios discussed above \cite{Renzini81-AA,Gratton04-ARAA,Ventura08-AAp,Denisenkov90-SAL,Jose98-ApJ,Iliadis02-ApJSS,Denissenkov14-MNRAS-442,Chamulak:2008-ApJ,Pignatari15-ApJL}, the strengths of resonances with $E_{\rm p}^{\rm res}$\,$<$\,400\,keV must be known.
Only one direct experiment is reported in the literature \cite{Goerres82-NPA}, and it shows only upper limits for the resonance strengths. Indirect data are also available \cite{Powers71-PRC,Hale01-PRC,Jenkins:2013fna}, but their interpretation relies on spin parity assignments or spectroscopic factor normalizations which are often uncertain. 
As a result, the mere existence of the resonances at $E_{\rm p}^{\rm res}$ = 71, 105 and 215\,keV is still under debate \cite{Powers71-PRC,Hale01-PRC}.

In 1999, the NACRE collaboration \cite{NACRE99-NPA} derived the $^{22}$Ne(p,$\gamma$)$^{23}$Na reaction rate from resonance strengths \cite{Goerres82-NPA, Goerres83-NPA,Meyer73-NPA} and a small direct capture component \cite{Goerres83-NPA}. A similar evaluation was performed by Hale {\it et al.} in 2001 \cite{Hale01-PRC}, updated by Iliadis {\it et al.} in 2010 \cite{Iliadis10-NPA841_31,Iliadis10-NPA841_251} and again in 2013 by the STARLIB group \cite{Sallaska13-ApJS}, including new indirect data \cite{Hale01-PRC}. Iliadis {\it et al.} used much lower upper limits than NACRE in several cases and excluded some debated resonances from consideration \cite{Iliadis10-NPA841_31,Iliadis10-NPA841_251,Sallaska13-ApJS}. As a result, there is up to a factor of 1000 difference in the total reaction rate between NACRE and STARLIB \cite{NACRE99-NPA,Sallaska13-ApJS}. The aim of the present work is to address this unsatisfactory situation with high-statistics, direct experimental data.

The measurements were carried out at the Laboratory for Underground Nuclear Astrophysics (LUNA) in the underground facility of the Italian National Institute for Nuclear Physics Gran Sasso National Laboratory, which offers an unprecedented sensitivity thanks to its low-background environment \cite{Costantini09-RPP,Broggini10-ARNPS}. Several nuclear reactions of astrophysical importance have been studied at very low energies at LUNA in recent years \cite{Bemmerer06-PRL,Scott12-PRL,Anders14-PRL}. The experimental setup (Refs.\,\cite{Cavanna:2014lia,Depalo15-Diss,Cavanna15-Diss} and Fig.\,\ref{fig:setup}) consists of a windowless gas target chamber filled with 1.5\,mbar $^{22}$Ne gas (isotopic enrichment 99.9\%, recirculated through a Monotorr II PS4-MT3-R-2 chemical getter) and two large high-purity germanium detectors, respectively at 55$^{\circ}$ (Ge55) and at 90$^{\circ}$ angle (Ge90) to the beam axis. Possible gas impurities by in-leaking air were periodically checked by the strong $E_{\rm p}^{\rm res}$\,=\,278\,keV $^{14}$N(p,$\gamma$)$^{15}$O resonance and always below 0.1\%.

The 70-300\,keV proton beam from the LUNA 400\,kV accelerator \cite{Formicola03-NIMA} (beam current 100-250\,$\mu$A) is collimated through a series of long, narrow apertures, then enters the target chamber, and is finally stopped on a copper beam calorimeter with constant temperature gradient. Ge90 and Ge55 are surrounded by a 4$\pi$ lead shield of 22\,-\,25\,cm thickness, and a 4\,cm inner copper liner for Ge55.
The $\gamma$-ray detection efficiency was measured (478\,keV\,$\leq$\,$E_{\gamma}$\,$\leq$\,1836\,keV) with calibrated radioactive sources ($^7$Be, $^{60}$Co, $^{88}$Y, $^{137}$Cs) and extended to higher energies (765\,keV\,$\leq$\,$E_{\gamma}$\,$\leq$\,6790\,keV) by the isotropic 1:1 photon cascades of the $^{14}$N(p,$\gamma$)$^{15}$O resonance at $E_{\rm p}^{\rm res}$\,=\,278\,keV.
          
As a first step, the energies of the resonances at $E_{\rm p}^{\rm res}$\,=\,271.6\,keV in $^{21}$Ne(p,$\gamma$)$^{22}$Na and 384.5\,keV in $^{20}$Ne(p,$\gamma$)$^{21}$Na reactions have been re-measured, relative to the accelerator energy calibration ($\Delta E_{\rm p}$ = 0.3\,keV \cite{Formicola03-NIMA}). Using 
neon gas of natural isotopic composition (90.48\% $^{20}$Ne, 0.27\% $^{21}$Ne, 9.25\% $^{22}$Ne), resonance energies of 271.5\,$\pm$\,1.0 and 384.5\,$\pm$\,0.5\,keV, respectively, were found, consistent with the literature \cite{Becker92-ZPA,Iliadis10-NPA841_251}. This confirms that accelerator energy calibration and energy loss in the gas are properly understood.

\begin{figure}[tb]
\includegraphics[width=\columnwidth]{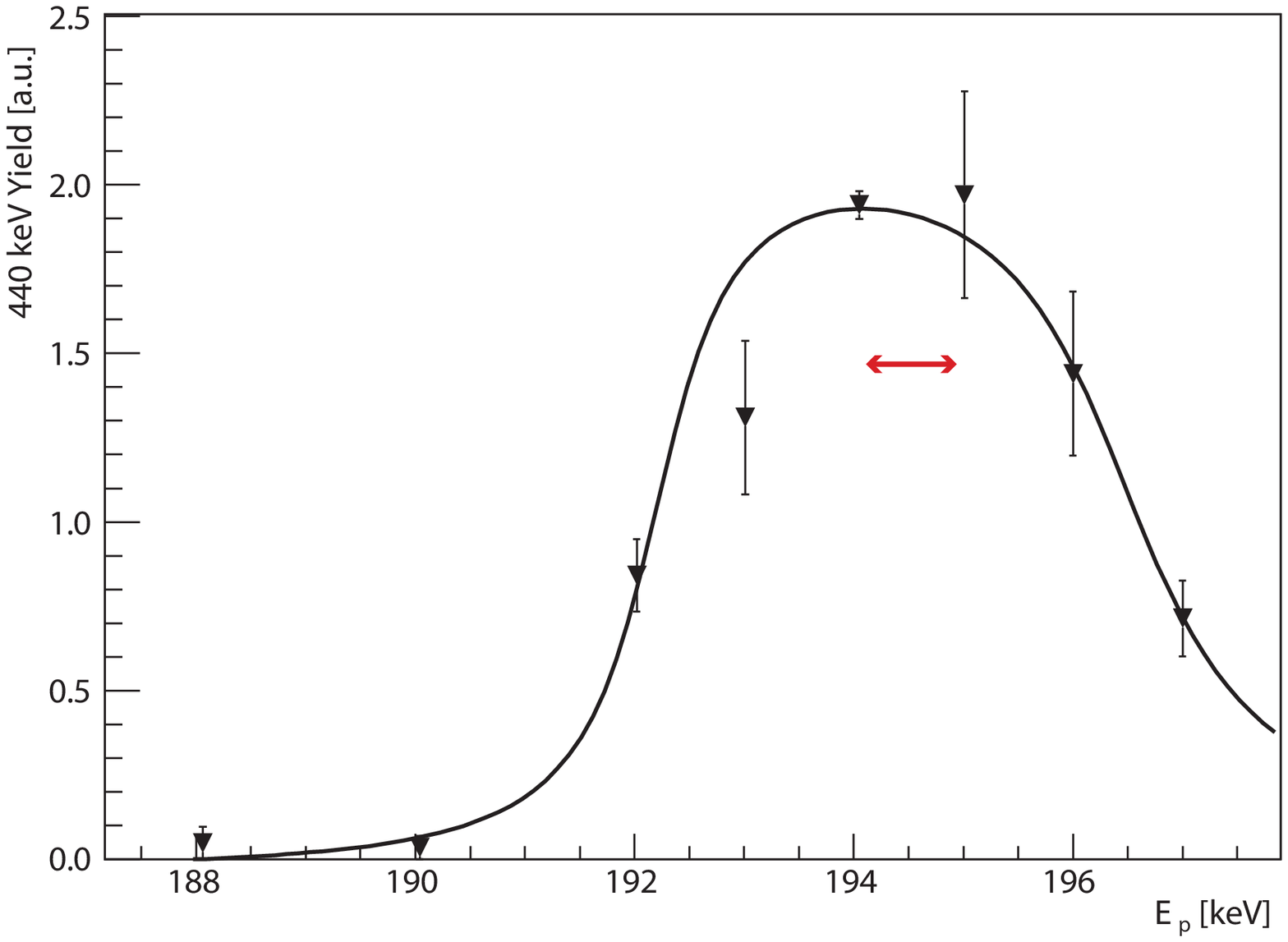}
\caption{Ge55 scan of the resonance at $E_{\rm p}^{\rm res}$\,=\,189.5\,keV. The maximum at $E_{\rm p}$\,=\,194.5\,keV corresponds to a resonance energy of (189.5\,$\pm$\,0.7)\,keV. The red arrow indicates the uncertainty on the resonance energy. The black line is just to guide the eye.}
\label{fig:scan}
\end{figure}

\begin{figure*}[tb]
\includegraphics[width=\textwidth]{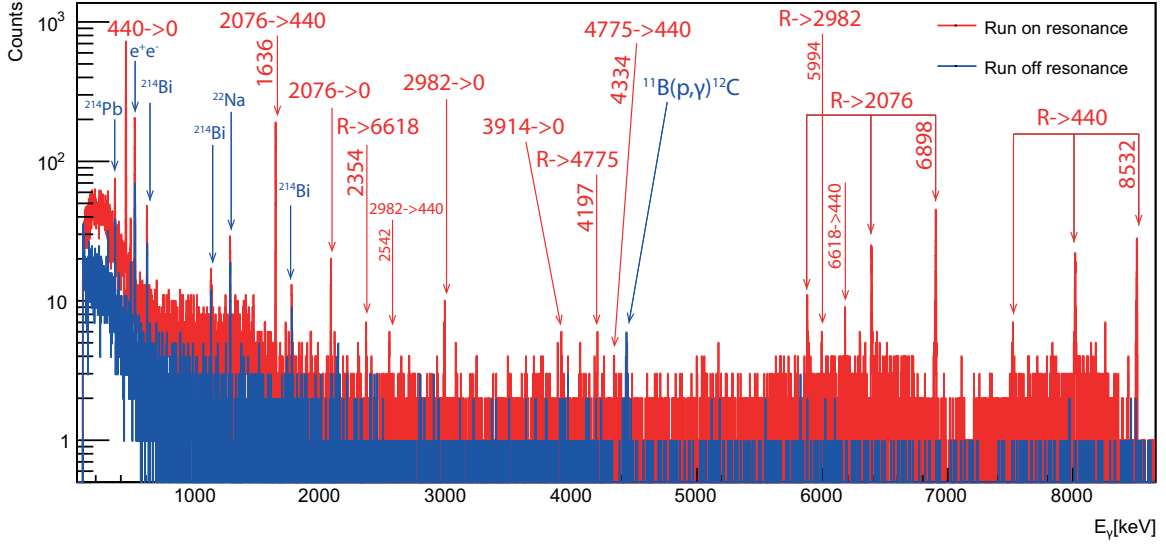}
\caption{(Color online) Spectra taken with Ge55 at $E_{\rm p}$\,=\,195\,keV, on the top of the 189.5\,keV resonance (red), and at $E_{\rm p}$\,=\,221\,keV, well outside the resonance profile, scaled by a factor of 0.99 for equal charge (blue). Red arrows show $^{22}$Ne(p,$\gamma$)$^{23}$Na transitions (``\,R\,'' denotes the resonance, numbers  the $^{23}$Na excitation energy in keV). Blue arrows show background lines.}
\label{fig:spectra}
\end{figure*}

Each of the suspected $^{22}$Ne(p,$\gamma$)$^{23}$Na resonances \citep{Goerres82-NPA,Iliadis10-NPA841_251} was first scanned with 3-9 beam energy steps of 1-2\,keV (Fig.\,\ref{fig:scan}). If a resonance was indeed detected, its energy was then obtained by matching the yield profile for the 440\,keV $\gamma$ ray (de-excitation of the first excited state in $^{23}$Na) with the efficiency profile taken with the $^7$Be source ($E_{\gamma}$\,=\,478\,keV). New $^{22}$Ne(p,$\gamma$)$^{23}$Na resonances were found at 156.2\,$\pm$\,0.7\,keV, 189.5\,$\pm$\,0.7\,keV and 259.7\,$\pm$\,0.6\,keV. The uncertainty includes 1.7\% error on the proton energy loss in neon gas ($\sim$\,0.5\,keV/cm) \cite{Ziegler10-NIMB}. The corresponding $^{23}$Na excitation energies are consistent with, but more precise than, the literature values \citep{Jenkins:2013fna,Iliadis10-NPA841_251} except for the resonance at 189.5\,keV. The reported \citep{Jenkins:2013fna} and adopted \citep{Iliadis10-NPA841_251} level energy of $E_{\rm x}$\,=\,8972\,keV is instead found to be 8975.3\,$\pm$\,0.7\,keV here.
After the scan, high-statistics runs (typical running time 58\,h on top of the $E_{\rm p}^{\rm res}$\,=\,156.2\,keV resonance) were performed at the maximum of the yield profile and, to determine the non-resonant yield, well outside the resonance profile (Fig.\,\ref{fig:scan}) at 9-53\,keV distance in an area without other suspected resonances and with low beam-induced background. The observed on-resonance spectra are dominated by the resonance under study, see Fig.\,\ref{fig:spectra} for typical Ge55 spectra. The non-resonant yield was always found to be consistent with zero for the resonances observed here. 

The resonance strength $\omega\gamma$ was then determined from the total yield, $Y_{\rm max}$, given by the sum of the primary transitions, (i.e. transitions from the resonance under study to a given state in $^{23}$Na) after correcting for detection efficiency:
\begin{equation}
\label{eq:wg_def}
\omega\gamma=\frac{2\,Y_{\rm max}\,\epsilon_{\rm R}}{\lambda^2_{\rm R}}\frac{m_{\rm t}}{m_{\rm t}+m_{\rm p}}
\end{equation}
where $\epsilon_{\rm R}$ is the effective stopping power in the laboratory system, $\lambda^2_{\rm R}$ is the squared de Broglie wavelength at the resonance energy, and $m_{\rm t}$ and $m_{\rm p}$ are the masses of target and projectile, respectively. The spin-parity of the resonances shown here is partly still under study \cite{Jenkins:2013fna}, therefore theoretical predictions of the angular distributions are fraught with uncertainty. For each resonance, the $\omega\gamma$ data obtained by separately analyzing Ge55 and Ge90 were found to be mutually consistent within 3-14\% statistical uncertainty. Isotropy was then assumed, and the weighted average of Ge55 and Ge90 was adopted (Table \ref{tab:wg_TRR}). The systematic uncertainty includes 3\% uncertainty on the $\gamma$ detection efficiency, 0.6\% on the effective gas density \citep{Cavanna:2014lia}, and 1\% on the beam current.

Sources of beam induced background include the $^{11}$B(p,$\gamma$)$^{12}$C, $^{12}$C(p,$\gamma$)$^{13}$N, and $^{19}$F(p,$\alpha\gamma$)$^{16}$O reactions \citep{Cavanna:2014lia}. The former two have only negligible impact. The latter affects spectra for $E_{\rm p}>$340\,keV \citep{Cavanna:2014lia}. Even given this background, for example in fig.\,\ref{fig:spectra} the continuum at 6-8\,MeV is still a factor of 50 lower then the no-beam background at the surface of the Earth \citep{Szucs10-EpjA}. 

The present resonance strength values are consistent with the previous direct upper limits \cite{Goerres82-NPA}, with the exception of the resonance at $E_{\rm p}^{\rm res}$\,=\,259.7\,keV which is slightly stronger than the previous upper limit (Table \ref{tab:wg_TRR}). Interesting discrepancies appear when comparing the present direct data with the previous indirect upper limits. For the resonances at 156.2\,keV and 259.7\,keV, the present strengths are a factor of 16 and 50, respectively, higher than the indirect upper limits \cite{Hale01-PRC,Iliadis10-NPA841_251}. For these two states, only limited angular distribution data were available in the indirect works \cite{Hale01-PRC}, hampering their interpretation. Indeed, new spin-parity values reported recently \cite{Jenkins:2013fna} are somewhat different from those in Ref.\,\cite{Hale01-PRC}. Together with possible normalization issues, this fact might explain the discrepancy with the present direct data.
In the present measurement there was no evidence of the suspected weak resonances at 71, 105, and 215\,keV. The profile likelihood method \cite{Rolke-2001NIMPA} has been used to derive new upper limits (90\% confidence level) from the present data for these cases. The new upper limits (Table\,\ref{tab:wg_TRR}) are much improved with respect to the literature. 
Resonances at $E_{\rm p}^{\rm res}$ = 290-400\,keV were not explored here, because the
previous direct \cite{Goerres82-NPA} and indirect \cite{Hale01-PRC} upper limits already show that they contribute only negligibly, $<$\,1\%, to the total thermonuclear reaction rate.
 
%

\begin{table*}[tb]
\footnotesize{
\caption{$^{22}$Ne(p,$\gamma$)$^{23}$Na resonance strengths from the literature and from the present work. The error bar on the strength measured in this work includes both statistical and systematic uncertainties.}
\begin{center}
\label{tab:wg_TRR}
\begin{tabular}{ c c | c | r | c | c | c  c }
\hline
\multicolumn{2}{c}{\textbf{Energy\,[keV]}} & \multicolumn{5}{c}{\textbf{Strength} $\omega\gamma$\,[eV]} \\
$E_{\rm p}^{\rm res}$ & $E_{\rm x}$ & \multicolumn{1}{c}{Previous direct} & \multicolumn{1}{c}{Previous indirect} & NACRE\,\cite{NACRE99-NPA} & Iliadis et al.\,\cite{Iliadis10-NPA841_251} & Present work & Present work \\
   &  & &  & adopted & adopted & direct & adopted \\
\hline
\hline
29 & 8822 & - & $\leq$\,2.6\,10$^{-25}$\,\cite{Iliadis10-NPA841_251} & - & $\leq$\,2.6\,10$^{-25}$ & - & $\leq$\,2.6\,10$^{-25}$ \\
37 & 8829 & - & (3.1\,$\pm$\,1.2)\,10$^{-15}$\,\cite{Iliadis10-NPA841_251} & (6.8\,$\pm$\,1.0)\,10$^{-15}$ & (3.1\,$\pm$\,1.2)\,10$^{-15}$ & - & (3.1\,$\pm$\,1.2)\,10$^{-15}$ \\ 
71 & 8862 & $\leq$\,3.2\,10$^{-6}$\,\cite{Goerres82-NPA} & $\leq$\,1.9\,10$^{-10}$\,\cite{Hale01-PRC} & $\leq$\,4.2\,10$^{-9}$ & - & $\leq$\,1.5\,10$^{-9}$ & $\leq$\,1.5\,10$^{-9}$\\
105 & 8895 & $\leq$\,0.6\,10$^{-6}$\,\cite{Goerres82-NPA} & $\leq$\,1.4\,10$^{-7}$\,\cite{Hale01-PRC} & $\leq$\,6.0\,10$^{-7}$ & - & $\leq$\,7.6\,10$^{-9}$ & $\leq$\,7.6\,10$^{-9}$ \\
156.2\,$\pm$\,0.7 & 8943.5\,$\pm$\,0.7 & $\leq$\,1.0\,10$^{-6}$\,\cite{Goerres82-NPA} & (9.2\,$\pm$\,3.7)\,10$^{-9}$\,\cite{Iliadis10-NPA841_251} & (6.5\,$\pm$\,1.9)\,10$^{-7}$ & (9.2\,$\pm$\,3.7)\,10$^{-9}$ & [1.48\,$\pm$\,0.10]\,10$^{-7}$ & [1.48\,$\pm$\,0.10]\,10$^{-7}$ \\
189.5\,$\pm$\,0.7 & 8975.3\,$\pm$\,0.7 & $\leq$\,2.6\,10$^{-6}$\,\cite{Goerres82-NPA} & $\leq$\,2.6\,10$^{-6}$
\,\cite{Hale01-PRC} & $\leq$\,2.6\,10$^{-6}$ & $\leq$\,2.6\,10$^{-6}$ & [1.87\,$\pm$\,0.06]\,10$^{-6}$ & [1.87\,$\pm$\,0.06]\,10$^{-6}$\\
215 & 9000 & $\leq$\,1.4\,10$^{-6}$\,\cite{Goerres82-NPA} & - & $\leq$\,1.4\,10$^{-6}$ & - & $\leq$\,2.8\,10$^{-8}$ & $\leq$\,2.8\,10$^{-8}$ \\
259.7\,$\pm$\,0.6 & 9042.4\,$\pm$\,0.6 & $\leq$\,2.6\,10$^{-6}$\,\cite{Goerres82-NPA} & $\leq$\,1.3\,10$^{-7}$\,\cite{Hale01-PRC} & $\leq$\,2.6\,10$^{-6}$ & $\leq$\,1.3\,10$^{-7}$ & [6.89\,$\pm$\,0.16]\,10$^{-6}$ & [6.89\,$\pm$\,0.16]\,10$^{-6}$ \\
291 & 9072 & $\leq$\,2.2\,10$^{-6}$\,\cite{Goerres82-NPA} & - & $\leq$\,2.2\,10$^{-6}$ & $\leq$\,2.2\,10$^{-6}$ & - & $\leq$\,2.2\,10$^{-6}$ \\
323 & 9103 & $\leq$\,2.2\,10$^{-6}$\,\cite{Goerres82-NPA} & - & $\leq$\,2.2\,10$^{-6}$ & $\leq$\,2.2\,10$^{-6}$ & - & $\leq$\,2.2\,10$^{-6}$ \\
334 & 9113 & $\leq$\,3.0\,10$^{-6}$\,\cite{Goerres82-NPA} & - & $\leq$\,3.0\,10$^{-6}$ & $\leq$\,3.0\,10$^{-6}$ & - & $\leq$\,3.0\,10$^{-6}$ \\
369 & 9147 & - & $\leq$\,6.0\,10$^{-4}$\,\cite{Hale01-PRC} & - & $\leq$\,6.0\,10$^{-4}$ & - & $\leq$\,6.0\,10$^{-4}$ \\
394 & 9171 & - & $\leq$\,6.0\,10$^{-4}$\,\cite{Hale01-PRC} & - & $\leq$\,6.0\,10$^{-4}$ & - & $\leq$\,6.0\,10$^{-4}$\\
\hline          
\end{tabular}
\end{center}}
\end{table*}%

Using the present new low-energy resonance strengths at novae and AGB star energies (Table\,\ref{tab:wg_TRR}), literature strengths \cite{Depalo15-PRC,Iliadis10-NPA841_251} at energies not explored here, and the previously assumed non-resonant S-factor of 62\,keV\,b \cite{Iliadis10-NPA841_251} (contributing $<$\,5\% to the rate), the thermonuclear reaction rate was calculated in the temperature range 0.02-1\,GK. The resonant contribution, in units of cm$^3$ mol$^{-1}$ s$^{-1}$, is given by \cite{NACRE99-NPA}: 
\begin{equation}
\label{eq:rate_tot_res}
N_{\rm A}\langle \sigma v\rangle_{\rm R}\,=\,\frac{1.5399\,10^{5}}{(\mu T_9)^{3/2}} \displaystyle\sum_i (\omega\gamma)_i e^{-11.605 E^{\rm res}_{\rm i}/T_9}
\end{equation}
where $N_{\rm A}$ denotes Avogadro's constant, $T_9$ is the temperature in units of GK, $\omega\gamma_{\rm i}$ is the resonance strength in eV (Table \ref{tab:wg_TRR}, last column), $\mu$ and $E^{\rm res}_i$ are the reduced mass (in amu) and the resonance energy in the center of mass (in MeV). In order to estimate the total rate and its uncertainty, a Monte Carlo method has been employed: (i) For each resonance, the two input parameters resonance energy and strength are each sampled from a gaussian probability density distribution, taking into account their values and uncertainties, from the present data, where available, and from the literature \cite{Iliadis10-NPA841_31} for the other cases; (ii) the reaction rate is calculated for a set of temperatures in the 0.02-1\,GK range; (iii) steps (i)-(ii) are repeated 10,000 times. For the resonances at 71, 105, and 215\,keV, where only upper limits exist, in step (ii) the resonance strength was sampled by separately sampling Poisson distributions for signal and background excluding unphysical negative values when taking the difference \cite{Rolke-2001NIMPA}. This procedure yields the probability density function and thus a median value and uncertainty of the reaction rate at each temperature. In order to be very conservative, for the lower side of the 1$\sigma$ error band the unobserved resonances at 71, 105, and 215\,keV were forced to zero, exactly as STARLIB \cite{Iliadis10-NPA841_31,Iliadis10-NPA841_251,Sallaska13-ApJS} did for these resonances.

\begin{figure}[tb]
\centering
\includegraphics[trim=0 0 6cm 0,clip,width=\columnwidth]{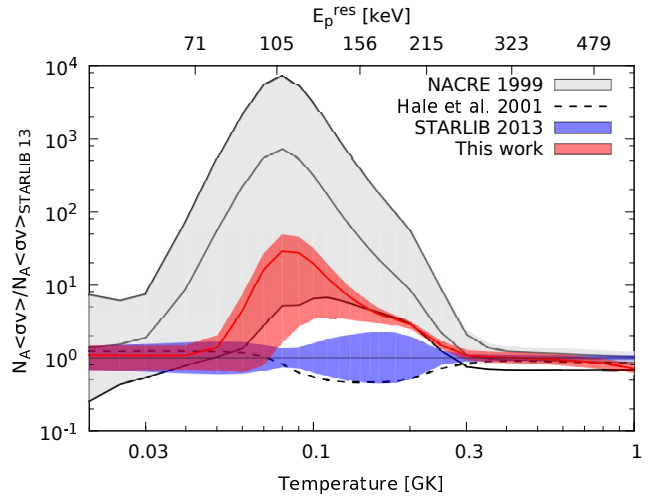}
\caption{Thermonuclear reaction rate of this work, NACRE\,\cite{NACRE99-NPA}, Hale  {\it et al.} \cite{Hale01-PRC}, and STARLIB\,\cite{Sallaska13-ApJS} normalized to STARLIB\,\cite{Sallaska13-ApJS}.}
\label{fig:TRR_LUNA-NACRE-Iliadis}
\end{figure}

The central value of the present new thermonuclear reaction rate lies between those of NACRE and STARLIB \cite{Sallaska13-ApJS} (Fig.\,\ref{fig:TRR_LUNA-NACRE-Iliadis}). It is consistent, within the previous significant error bars, with NACRE. 
The new 1$\sigma$ lower limit is above the previous upper limit by STARLIB \cite{Sallaska13-ApJS} for 0.08\,GK\,$<$\,$T$\,$<$\,0.25\,GK, mainly because of the newly observed resonances at 156.2\,keV, 189.5\,keV, and 259.7\,keV. 
The larger error bar at low temperatures, 0.05-0.1\,GK, is explained by the different treatment of the 71, 105, and 215\,keV resonances. In the present work, they are set to zero for the lower error bar, but the Monte Carlo approach includes them for the median value and upper error bar. Previously, they were excluded and did not contribute to the uncertainty \cite{Iliadis10-NPA841_31,Iliadis10-NPA841_251,Sallaska13-ApJS}. 

In order to illustrate the impact of the new rate on $^{23}$Na and the Na-O anticorrelation, previous calculated $^{23}$Na yields for intermediate mass AGB stars \cite{Izzard07-AeA} are used here. That calculation \cite{Izzard07-AeA} used the Hale {\it et al.} \cite{Hale01-PRC} reaction rate as control value; the present rate is a factor of 10-30 higher at HBB temperatures. The new rate leads to an enhancement of the predicted $^{23}$Na yield by a factor of three for an AGB star model of 5\,M$_{\odot}$ and initial metallicity $Z$ = 0.008 \cite{Izzard07-AeA}. Other HBB simulations \cite{Ventura08-AAp} have reported that the most oxygen-poor ejecta are also sodium-rich, again based on the Hale {\it et al.} \cite{Hale01-PRC} rate. These ejecta \cite{Ventura08-AAp} will become even more sodium rich, by about a factor of three, based on the present, new rate.

Summarizing, the $^{22}$Ne(p,$\gamma$)$^{23}$Na reaction has been studied by underground in-beam $\gamma$ spectroscopy using a windowless gas target. Three low-energy resonances have been observed for the first time, leading to a significant increase of the thermonuclear reaction rate at temperatures 0.08\,$\leq$\,$T$\,$\leq$\,0.3\,GK relevant to AGB stars, hot bottom burning, and novae. Significantly reduced upper limits were obtained for three more resonances; for further progress here, a new experiment with even higher luminosity is needed.

\begin{acknowledgments}
Financial support by INFN, DFG (BE 4100-2/1), NAVI (HGF VH-VI-417), OTKA (K101328), and a DAAD fellowship at HZDR for F.C. are gratefully acknowledged. 
\end{acknowledgments}


\begin{thebibliography}{10}

\bibitem{Carretta:2011-AA}
E.~{Carretta}, S.~{Lucatello}, R.~G. {Gratton}, A.~{Bragaglia}, and
  V.~{D'Orazi},
\newblock Astron.~Astrophys. {\bf 533}, A69 (2011).

\bibitem{Lind:2011-AA}
{Lind, K.}, {Charbonnel, C.}, {Decressin, T.}, {Primas, F.}, {Grundahl, F.},
  and {Asplund, M.},
\newblock \aap {\bf 527}, A148 (2011).

\bibitem{Yong:2003-AA}
{D. Yong}, {F. Grundahl}, {D. L. Lambert}, {P. E. Nissen}, and {M. D.
  Shetrone},
\newblock \aap {\bf 402}, 985 (2003).

\bibitem{Gratton12-AAR}
R.~G. {Gratton}, E.~{Carretta}, and A.~{Bragaglia},
\newblock \aar {\bf 20}, 50 (2012).

\bibitem{Cordero15-AJ}
M.~J. Cordero, C.~A. Pilachowski, C.~I. Johnson, and E.~Vesperini,
\newblock \apj {\bf 800}, 3 (2015).

\bibitem{Gratton04-ARAA}
R.~{Gratton}, C.~{Sneden}, and E.~{Carretta},
\newblock \araa {\bf 42}, 385 (2004).

\bibitem{Carretta09-AA}
E.~{Carretta} {\em et~al.},
\newblock Astron.~Astrophys. {\bf 505}, 117 (2009).

\bibitem{Johnson12-ApJL}
C.~I. {Johnson} and C.~A. {Pilachowski},
\newblock \apjl {\bf 754}, L38 (2012).

\bibitem{Renzini81-AA}
A.~{Renzini} and M.~{Voli},
\newblock Astron.~Astrophys. {\bf 94}, 175 (1981).

\bibitem{Ventura08-AAp}
P.~{Ventura} and F.~{D'Antona},
\newblock \aap {\bf 479}, 805 (2008).

\bibitem{deMink09-aap}
S.~E. {de Mink}, O.~R. {Pols}, N.~{Langer}, and R.~G. {Izzard},
\newblock \aap {\bf 507}, L1 (2009).

\bibitem{Decressin07-aap}
T.~{Decressin}, G.~{Meynet}, C.~{Charbonnel}, N.~{Prantzos}, and
  S.~{Ekstr{\"o}m},
\newblock \aap {\bf 464}, 1029 (2007).

\bibitem{Denissenkov14-MNRAS}
P.~A. {Denissenkov} and F.~D.~A. {Hartwick},
\newblock \mnras {\bf 437}, L21 (2014).

\bibitem{Denisenkov90-SAL}
P.~A. {Denisenkov} and S.~N. {Denisenkova},
\newblock Soviet Astronomy Letters {\bf 16}, 275 (1990).

\bibitem{Strieder12-PLB}
F.~{Strieder} {\em et~al.},
\newblock Phys.~Lett.~B {\bf 707}, 60 (2012).

\bibitem{Scott12-PRL}
D.~A. {Scott} {\em et~al.},
\newblock Phys.~Rev.~Lett. {\bf 109}, 202501 (2012).

\bibitem{Cavanna:2014lia}
F.~Cavanna {\em et~al.},
\newblock \epja {\bf 50}, 179 (2014).

\bibitem{Depalo15-Diss}
R.~Depalo,
\newblock {\em {The neon-sodium cycle: Study of the
  $\rm^{22}Ne(p,\gamma)^{23}Na$ reaction at astrophysical energies}},
\newblock PhD thesis, Universit\`a di Padova, 2015.

\bibitem{Cavanna15-Diss}
F.~Cavanna,
\newblock {\em {A direct measurement of the $\rm^{22}Ne(p,\gamma)^{23}Na$
  reaction down to the energies of astrophysical interest}},
\newblock PhD thesis, Universit\`a di Genova, 2015.

\bibitem{Jose98-ApJ}
J.~Jos{\'e} and M.~Hernanz,
\newblock \apj {\bf 494}, 680 (1998).

\bibitem{Sallaska10-PRL}
A.~L. Sallaska {\em et~al.},
\newblock Phys. Rev. Lett. {\bf 105}, 152501 (2010).

\bibitem{Iliadis02-ApJSS}
C.~Iliadis, A.~Champagne, J.~Jos{\'e}, S.~Starrfield, and P.~Tupper,
\newblock Astrophys.~J.~Suppl.~Ser. {\bf 142}, 105 (2002).

\bibitem{Denissenkov14-MNRAS-442}
P.~A. {Denissenkov} {\em et~al.},
\newblock \mnras {\bf 442}, 2058 (2014).

\bibitem{Chamulak:2008-ApJ}
D.~A. Chamulak, E.~F. Brown, F.~X. Timmes, and K.~Dupczak,
\newblock The Astrophysical Journal {\bf 677}, 160 (2008).

\bibitem{Pignatari15-ApJL}
M.~Pignatari {\em et~al.},
\newblock \apjl {\bf 808}, L43 (2015).

\bibitem{Depalo15-PRC}
R.~{Depalo} {\em et~al.},
\newblock Phys.~Rev.~C {\bf 92}, 045807 (2015).

\bibitem{Goerres82-NPA}
J.~{G\"orres}, C.~{Rolfs}, P.~{Schmalbrock}, H.~P. {Trautvetter}, and
  J.~{Keinonen},
\newblock Nucl.~Phys.~A {\bf 385}, 57 (1982).

\bibitem{Powers71-PRC}
J.~R. {Powers}, H.~T. {Fortune}, R.~{Middleton}, and O.~{Hansen},
\newblock Phys.~Rev.~C {\bf 4}, 2030 (1971).

\bibitem{Hale01-PRC}
S.~E. {Hale}, A.~E. {Champagne}, C.~{Iliadis}, V.~Y. {Hansper}, D.~C. {Powell},
  and J.~C. {Blackmon},
\newblock Phys.~Rev.~C {\bf 65}, 015801 (2001).

\bibitem{Jenkins:2013fna}
D.~Jenkins {\em et~al.},
\newblock \prc {\bf 87}, 064301 (2013).

\bibitem{NACRE99-NPA}
C.~{Angulo} {\em et~al.},
\newblock Nucl.~Phys.~A {\bf 656}, 3 (1999).

\bibitem{Goerres83-NPA}
J.~{G{\"o}rres} {\em et~al.},
\newblock Nucl.~Phys.~A {\bf 408}, 372 (1983).

\bibitem{Meyer73-NPA}
M.~Meyer and J.~Smit,
\newblock \npa {\bf 205}, 177  (1973).

\bibitem{Iliadis10-NPA841_31}
C.~{Iliadis}, R.~{Longland}, A.~E. {Champagne}, A.~{Coc}, and R.~{Fitzgerald},
\newblock Nucl.~Phys.~A {\bf 841}, 31 (2010).

\bibitem{Iliadis10-NPA841_251}
C.~{Iliadis}, R.~{Longland}, A.~E. {Champagne}, and A.~{Coc},
\newblock Nucl.~Phys.~A {\bf 841}, 251 (2010).

\bibitem{Sallaska13-ApJS}
A.~L. {Sallaska}, C.~{Iliadis}, A.~E. {Champange}, S.~{Goriely},
  S.~{Starrfield}, and F.~X. {Timmes},
\newblock \apjs {\bf 207}, 18 (2013).

\bibitem{Costantini09-RPP}
H.~{Costantini}, A.~Formicola, G.~Imbriani, M.~Junker, C.~Rolfs, and
  F.~Strieder,
\newblock Rep.~Prog.~Phys. {\bf 72}, 086301 (2009).

\bibitem{Broggini10-ARNPS}
C.~Broggini, D.~Bemmerer, A.~Guglielmetti, and R.~Menegazzo,
\newblock Annu. Rev. Nucl. Part. Sci. {\bf 60}, 53 (2010).

\bibitem{Bemmerer06-PRL}
D.~{Bemmerer} {\em et~al.},
\newblock Phys.~Rev.~Lett. {\bf 97}, 122502 (2006).

\bibitem{Anders14-PRL}
M.~Anders {\em et~al.},
\newblock Phys. Rev. Lett. {\bf 113}, 042501 (2014).

\bibitem{Formicola03-NIMA}
A.~{Formicola} {\em et~al.},
\newblock Nucl.~Inst.~Meth.~A {\bf 507}, 609 (2003).

\bibitem{Becker92-ZPA}
H.~W. {Becker} {\em et~al.},
\newblock Z.~Phys.~A {\bf 343}, 361 (1992).

\bibitem{Ziegler10-NIMB}
J.~F. Ziegler, M.~D. Ziegler, and J.~P. Biersack,
\newblock Nucl.~Inst.~Meth.~B {\bf 268}, 1818 (2010).

\bibitem{Szucs10-EpjA}
T.~{Sz{\"u}cs} {\em et~al.},
\newblock Eur.~Phys.~J.~A {\bf 44}, 513 (2010).

\bibitem{Rolke-2001NIMPA}
W.~A. {Rolke} and A.~M. {L{\'o}pez},
\newblock \nima {\bf 458}, 745 (2001).

\bibitem{Izzard07-AeA}
R.~G. {Izzard}, M.~{Lugaro}, A.~I. {Karakas}, C.~{Iliadis}, and M.~{van Raai},
\newblock \aap {\bf 466}, 641 (2007).

\end{thebibliography}
\end{document}